\documentclass[twoside]{ilcws07}
\usepackage[latin1]{inputenc}
\usepackage[dvips]{graphicx,epsfig,color}
\usepackage{wrapfig,rotating}
\usepackage{amssymb,amsmath,array}

\begin{document}

\title{
On-shell renormalisation constants including two different nonzero masses} 

\author{S. Bekavac, D. Seidel
\vspace{.3cm}\\
Institut f{\"u}r Theoretische Teilchenphysik,
Universit{\"a}t Karlsruhe (TH)\\
76128 Karlsruhe, Germany
}

\maketitle

\begin{abstract}
We present results for the effect of a second massive quark in the relation between
the pole and the minimal subtracted quark mass at the three loop level. We also
consider the analogous effect for the wave function renormalisation constant. Some
technical details of the calculation are given. Our result is phenomenologically
relevant for the bottom quark including virtual charm effects.
\end{abstract}

\section{Introduction}

Quark masses are fundamental parameters of the Standard Model (SM) and thus it is
desirable to determine their numerical values with the highest possible precision. In
order to do so it is necessary to fix a renormalisation scheme which defines the
quark mass. Often physical observables are expressed through some ``short distance''
mass~\cite{SDM} to obtain predictions which are free of the renormalon problem. To do
so, one frequently needs the relation between the on-shell and the $\overline{\rm
MS}$ mass. Many authors contributed to the latter
task~\cite{MSOS,Hoang:2000fm,Marquard:2007uj}. In this contribution we present the
recently published calculation~\cite{Bekavac:2007tk}, where we have included the
effect of a second nonzero quark mass to the relation between the quarks in those two
schemes at the three loop level. After having reduced the problem to the calculation
of master integrals we use two different ways to solve them, namely the differential
equation and the Mellin-Barnes method. From the phenomenological point of view this
result is important for the bottom-quark including effects from virtual charm-quarks.
As a byproduct we also obtain the corresponding contribution to the wave function
renormalisation constant.

\section{Renormalisation constants}

Introducing the decomposition of the quark self-energy
\begin{eqnarray}
  \Sigma(q,m_q) &=&
  m_q\, \Sigma_1(q^2,m_q) + (q\!\!\!/\,\, - m_q)\, \Sigma_2(q^2,m_q)\,,
\end{eqnarray}
we can express the renormalisation constants, which are defined through
\begin{eqnarray}
  m_{q,0} = Z_m^{\rm OS}\, M_q\,,\qquad
  \psi_0 = \sqrt{Z_2^{\rm OS}}\, \psi\,,
\end{eqnarray}
by~\cite{Melnikov:2000zc,Marquard:2007uj}
\begin{eqnarray}
  Z_m^{\rm OS} &=& 1 + \Sigma_1(M_q^2,M_q)\,,
 \\
  \left( Z_2^{\rm OS} \right)^{-1} &=& 1 + 2M_q^2
  \frac{\partial}{\partial q^2} \Sigma_1(q^2,M_q) \Big|_{q^2 = M_q^2} +
  \Sigma_2(M_q^2,M_q) \,.
\end{eqnarray}
$\psi$ is the quark field renormalised in the on-shell scheme with mass $m_q$, $M_q$
is the on-shell mass and bare quantities are denoted by a subscript 0. Thus, to
obtain $Z_m^{\rm OS}$ one only needs to calculate $\Sigma_1$ for $q^2 = M_q^2$. To
calculate $Z_2^{\rm OS}$, one has to compute the first derivative of the self-energy
diagrams. The mass renormalisation is taken into account iteratively by calculating
one- and two-loop diagrams with zero-momentum insertions.

In the case of the mass renormalisation it is convenient to consider the ratio
between the on-shell and $\overline{\rm MS}$ renormalisation constants
\begin{eqnarray}
  z_m &=& \frac{Z_m^{\rm OS}}{Z_m^{\overline{\rm MS}}} \,\,=\,\,  
  \frac{m_q(\mu)}{M_q}
\end{eqnarray}
which is finite. Here we have introduced the renormalisation dependent
$\overline{\rm MS}$-mass $m_q(\mu)$.

In contrast to $Z_m^{\rm OS}$ the wave function renormalisation constant contains
next to ultraviolet also infrared divergences. Thus it is not possible to construct a
finite quantity by considering the ratio between the on-shell and $\overline{\rm MS}$
renormalisation constant.

\section{Computational techniques}

\begin{figure}[t]
  \begin{center}
    \epsfxsize=1\textwidth
    \epsffile{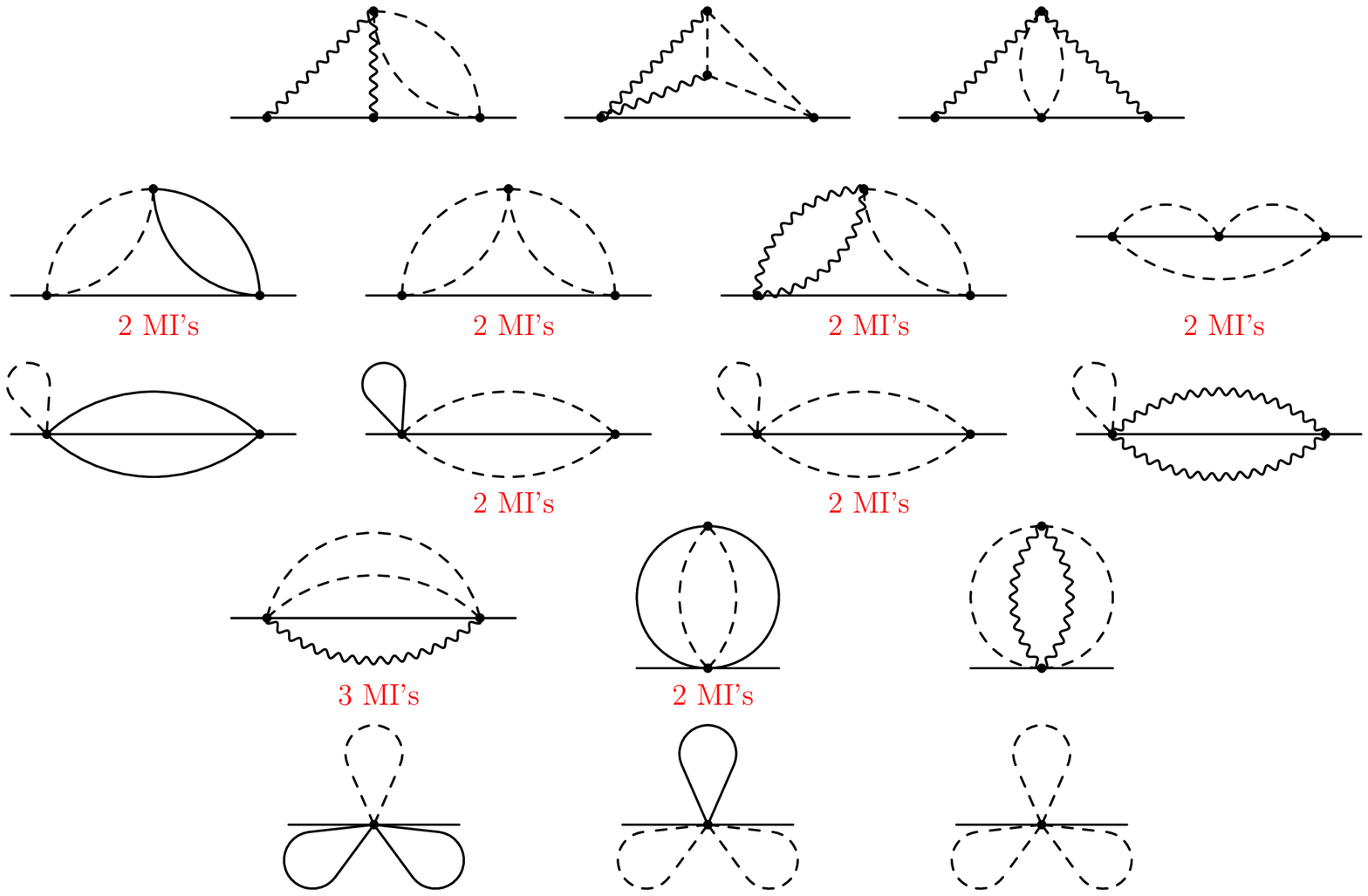}
  \end{center}
  \vspace*{-0em}
  \caption{
    \label{fig::MI} Master integrals. The solid/wavy lines correspond to particles
      with mass $M_q/0$, the dashed lines denote the quark with the second nonzero
      mass. }
\end{figure}

In order to compute the on shell self energy we use {\tt
QGRAF}~\cite{Nogueira:1991ex} to generate the feynman diagrams and the various
topologies are identified with the help of {\tt q2e} and {\tt
exp}~\cite{Harlander:1997zb,Seidensticker:1999bb}. In a next step the reduction of
the various functions to so-called master integrals (MI's) has to be achieved. For
this step we use the so-called Laporta method~\cite{Laporta:1996mq,Laporta:2001dd}
which reduces the three-loop integrals to 26 MI's. We use the implementation of
Laporta's algorithm in the program {\tt Crusher}~\cite{PMDS}. It is written in {\tt
C++} and uses {\tt GiNaC}~\cite{Bauer:2000cp} for simple manipulations like taking
derivatives of polynomial quantities. In the practical implementation of the Laporta
algorithm one of the most time-consuming operations is the simplification of the
coefficients appearing in front of the individual integrals. This task is performed
with the help of {\tt Fermat}~\cite{fermat} where a special interface has been used
(see Ref.~\cite{Tentyukov:2006ys}). The main features of the implementation are the
automated generation of the integration-by-parts (IBP)
identities~\cite{Chetyrkin:1981qh}, a complete symmetrisation of the diagrams and the
possibility to use multiprocessor environments.

In Fig.~\ref{fig::MI} a graphical representation of the master integrals can be
found. As indicated in the figure, many topologies contain more than one master
integral. We have chosen two independent ways to compute the $\varepsilon$-expansion
of the master integrals. The first one relies on the differential equation
method~\cite{Kotikov:1990kg}. With this method we were able to evaluate all but four
master integrals in analytic form. With the help of our second method, based on the
Mellin-Barnes technique (see, e.g., Ref.~\cite{Smirnov:2004ym}) we can get numerical
results for all master integrals. Here we have used the {\tt Mathematica} package
{\tt MB.m}~\cite{Czakon:2005rk}.

\subsection{Differential-Equation-Method}

First we set up differential equations in $z=M_f/M_q$, where $M_f$ is the second
nonzero quark mass, for each of the individual integrals. Each equation will contain
the master integral itself and integrals belonging to the corresponding topology. The
latter can again be reduced to the set of master integrals. For all the topologies
which contain only one master integral (e.g. all six-propagator integrals in
Fig.~\ref{fig::MI}) this gives a ``simple`` equation whereas for the other
topologies we get two or three coupled equations, respectively. In the next step we
expand the differential equations in $\varepsilon$. Choosing an appropriate basis for
the master integrals, all equations decouple order by order in $\varepsilon$. We can
now solve for all integrals by repeated use of Euler's variation of the constant
method. The initial conditions we need are all known from the
literature~\cite{Argeri:2002wz,Mastrolia:2002tv, Marquard:2006qi,Melnikov:2000zc}.

We were able to get analytical results for all master integrals in terms of Harmonic
Polylogarithms (HPL's~\cite{Remiddi:1999ew}) up to order $\varepsilon^{-1}$.
Unfortunately we were not yet able to get analytical results in higher orders in
$\varepsilon$ for the four integrals belonging to the two topologies depicted as the
leftmost ones in the second row of Fig.~\ref{fig::MI}. For all other integrals we
provide analytical results up to the order we need in our calculation.
To calculate the MI's in terms of HPL's it is necessary that the (pseudo)poles in the
corresponding differential equations are all of the form $1/z$, $1/(z+1)$ and
$1/(z-1)$, were these poles can occur up to arbitrary order. This is the case in all
topologies except for the ones mentioned above. We have not found a proper change of
variables to bring the differential equation into this form. As a consequence we only
managed to integrate the integrals in these topologies up to the order
$\varepsilon^{-1}$. The remaining parts can e.g. be integrated numerically with
Mathematica.

To evaluate our results numerically and for general algebraic manipulations of
terms involving HPL's we use the Mathematica package
HPL~\cite{Maitre:2005uu,Maitre:2007kp}.

\subsection{Mellin-Barnes-Method}

The Mellin-Barnes method as a tool for the evaluation of Feynman integrals has become
very popular in the recent years. The basic formula is \cite{Smirnov:2004ym}
\begin{equation}
\frac{1}{(X+Y)^\lambda} = \frac{1}{\Gamma(\lambda)} \frac{1}{2 \pi i} \int_{-i
\infty}^{i \infty} dz\, \Gamma(1+z) \Gamma(-z) \frac{Y^z}{X^{\lambda+z}},
\label{eq::mb-basic}
\end{equation}
which transforms a propagator like term into a complex contour integral. A common
recipe to evaluate Feynman integrals is the following: First one introduces Feynman
parameters for a loop variable. Then one can perform the corresponding momentum
integration. After that one applies formula (\ref{eq::mb-basic}) to the denominators
containing the Feynman parameters. Finally the Feynman parameters can be integrated
yielding the Mellin-Barnes representation of the original integral. This procedure 
has recently been automatised \cite{Gluza:2007rt}. 

The Mellin-Barnes integration is to be performed along a contour which reaches from
$-\infty$ to $\infty$ on the imaginary axis with indentations such that the poles of
$\Gamma(\ldots + z)$ and those of $\Gamma(\ldots - z)$ are separated by the contour.

Mellin-Barnes integrals usually have singularities for certain values of their
parameters. If there are for example Gamma functions of the form
$\Gamma(\varepsilon+z)\,\Gamma(-z)$, it is not possible to find an appropriate
integration contour when $\varepsilon \to 0$. The integral is therefore singular in
$\varepsilon$ and this is how UV poles of Feynman integrals manifest themselves in
their Mellin-Barnes representation. One has thus to regularise the integral, that is,
identify the  $\varepsilon$ poles. This can be done by shifting the integration
contour using the residue theorem. Prescriptions to do so have been given in Refs.
\cite{Smirnov:1999gc} and \cite{Tausk:1999vh}, the latter has been automatised, see
Refs. \cite{Anastasiou:2005cb, Czakon:2005rk}.

Finally the regularised integrals can be expanded in $\varepsilon$ and evaluated by
numerical integration, which is also implemented in the package {\ttfamily MB.m}
\cite{Czakon:2005rk}, or by application of the residue theorem and summing up the
residues. Depending on the dimension of the integrals and the complexity of the
integrand this can be done numerically or analytically.

To calculate the master integrals for this work we first simplified the Mellin-Barnes
integrals by inserting the representations of known subtopologies.  The
regularisation has been done partly using {\ttfamily MB.m} and the approach of Ref.
\cite{Smirnov:1999gc}. One- and two-dimensional MB-integrals were calculated via
their sum representation, higher dimensional integrals using {\ttfamily MB.m}.

The 4-line integrals can all be represented as Mellin-Barnes-integrals of maximal
dimension 1, which can be evaluated as single sums. For the 5 line master integrals
we find representations of dimension 2 to 5. The integrals with 6 lines have 3 to 5
dimensional representations. We find good agreement with the results obtained from
the differential equation method.

Inserting the results for the master integrals into the final result we observe large
numerical cancellations near $M_f=0$ between the contributions originating from
different master integrals. On the other hand, the expansion for $M_f/M_q\ll1$
converges very fast, which is relevant for $M_f=m_c$ and $M_q=m_b$. For this reason
we decided to derive an expansion of our result including eighth order terms. The
coefficients that could not be obtained analytically can be evaluated numerically
from their Mellin-Barnes-representation, which is in our case at most
two-dimensional.

\section{Results and applications}

As an application of our result we want to discuss the charm quark effects in the
relations between the pole, the $\overline{\rm MS}$ and the $1S$ quark mass. For
illustration we use $m_b(m_b)=4.2$~GeV, $m_c(m_c)=1.3$~GeV, $\mu=m_b$ and
$\alpha_s^{(4)}(m_b)=0.2247$. The relation between the on-shell and the
$\overline{\rm MS}$ mass leads to
\begin{eqnarray}
  M_b &=& \left[ 4.2 + 0.401 + \left(0.199 + 0.0094\Big|_{m_c}\right)
  + \left(0.145 + 0.0182\Big|_{m_c}\right) \right] \mbox{GeV}
  \,,
  \label{eq::mosmms}
\end{eqnarray}
where the tree-level, one-, two- and three-loop results are shown separately. The
contributions from the charm quark mass which vanish for $m_c\to 0$ are marked by a
subscript $m_c$. One observes that the higher order contributions are significant. In
particular, the two-loop charm quark effects amount to 9~MeV and the three-loop ones
to 18~MeV. A similar bad convergence is observed in the relation between the $1S$
mass~\cite{Hoang:1998nz} $M_b^{1S}$ and the pole mass $M_b$. For $M_b=4.8$~GeV,
$m_c(m_c)=1.3$~GeV, $\mu=M_b$ and $\alpha_s^{(4)}(M_b)=0.2150$ it is given by
\begin{eqnarray}
  M_b^{1S} &=& \left[4.8 - 0.049 - \left(0.073 + 0.0041\Big|_{m_c}\right)
  - \left(0.098 + 0.0112\Big|_{m_c}\right)\right] \mbox{GeV}
  \,.
  \label{eq::m1smos}
\end{eqnarray}
However, the relation between the $1S$ and the $\overline{\rm MS}$ quark mass is much
better behaved as can be seen in the following example where we have chosen
$M_b^{1S}=4.69$~GeV, $m_c(m_c)=1.3$~GeV, $\mu=M_b^{1S}$ and
$\alpha_s^{(4)}(M_b^{1S})=0.2167$
\begin{eqnarray}
  m_b &=& \left[ 4.69 - 0.382 - \left(0.098 + 0.0047\Big|_{m_c}\right)
  - \left(0.030 + 0.0051\Big|_{m_c}\right) \right] \mbox{GeV}
  \,.
  \label{eq::mmsm1s}
\end{eqnarray}
The two-loop charm effects amount to only 4.7~MeV and three-loop ones to 5.1~MeV. We
want to mention that in case only the linear approximation~\cite{Hoang:2000fm} of the
charm quark mass effects is used the corresponding three-loop results in
Eqs.~(\ref{eq::mosmms}) and~(\ref{eq::mmsm1s}) read $0.0167$ and $0.0037$,
respectively.

\section*{Acknowledgements}

We would like to thank Andrey Grozin and Matthias Steinhauser for a fruitful
collaboration on this subject. This work was supported the DFG through SFB/TR~9.

\begin{footnotesize}

\end{footnotesize}

\end{document}